\documentclass[twocolumn,twoside,slac_two]{revtex4}
\usepackage{graphicx}
\usepackage{fancyhdr}
\newcommand{\Bs}       {\ensuremath{B_{s}^{0}}}
\newcommand{\barBs}    {\ensuremath{\bar{B}_{s}^{0}}}
\newcommand{\Bd}       {\ensuremath{B_{d}^{0}}}

\newcommand{\Dsm}      {\ensuremath{D_{s}^{-}}}

\newcommand{\aonep}    {\ensuremath{a_{1}^{+}}}

\newcommand{\pip}      {\ensuremath{\pi^{+}}}

\newcommand{\ifb}  {\ensuremath{{\mathrm{fb}}^{-1}}}

\newcommand{\ips}  {\ensuremath{{\mathrm{ps}}^{-1}}}

\newcommand{\dms}  {\ensuremath{\Delta m_s}}
\newcommand{\dmd}  {\ensuremath{\Delta m_d}}
\newcommand{\dgs}  {\ensuremath{\Delta \Gamma_s}}
\newcommand{\dgsg} {\ensuremath{\Delta \Gamma_s /\Gamma_s}}

\newcommand{\pT}   {\ensuremath{p_\mathrm{T}}}

\begin{document}

\title{Prospects of the measurement of \boldmath{$B_{s}^{0}$} oscillations with the ATLAS detector at LHC }

%

\author{B.~Epp, E.~Kneringer, H.~Duer, P.~Jussel}
\affiliation{Institute of Astro- and Particle Physics, University of Innsbruck, Austria}

\begin{abstract}
An estimation of the sensitivity to measure
$\Bs$-$\barBs$ oscillations with the ATLAS detector
is given for the detector
geometry of ``initial layout''. The $\dms$ reach is derived from
unbinned maximum likelihood amplitude fits using $\Bs$ events
generated with a simplified Monte Carlo method.
\end{abstract}

\maketitle

\thispagestyle{fancy}

\section{\label{intro}Introduction}
The observed $\Bs$ and $\barBs$ particles are linear combinations of
the two mass eigenstates with masses $m_H$ and $m_L$ and a mass
difference of $\dms$. Transitions between the two flavor eigenstates
are allowed due to non--conservation of flavor in weak--current
interactions and will occur with a frequency proportional to \dms.
Together with the mass difference $\dmd$ of the $\Bd$~system, which
has already been measured with high accuracy ($\dmd = 0.502 \pm 0.007
\ips$\cite{PDG2004}), the measurement of $\dms$ is an important
ingredient for the precise determination of the side $|V_{td}|$ of the
CKM unitarity triangle.  The direct determination of $V_{td}$ and
$V_{ts}$ from \dmd\ and \dms\ is hampered by hadronic uncertainties.
These uncertainties partially cancel in the ratio of mass differences
\begin{eqnarray}
  \frac{\dms}{\dmd} = \xi^2 \;\frac{M_{\Bs}}{M_{\Bd}} \;
  \left | \frac {V_{ts}}{V_{td}} \right |^2 \;
\end{eqnarray}
Using the experimentally-measured masses $M_B$ and a value for the factor
$\xi$, which can be computed in lattice QCD, the constraint from the ratio
$\dms/\dmd$ is more effective in limiting the position of the apex of
the unitarity triangle than the value obtained by \dmd\ measurements
alone. \\
$\Bs$~oscillations have been observed recently at the Fermilab
Tevatron collider.  Whereas the D\O~collaboration is reporting a
two-sided bound $17 < \dms < 21 \ips$ at 90\%~CL~\cite{D0}, CDF
presents the first measurement of the $\Bs$-$\barBs$ oscillation
frequency finding a signal for
$\dms = 17.33\; ^{+0.42}_{-0.21}$\,(stat) $\pm 0.07$\,(sys)
at 95\%~CL~\cite{CDF}.  Both results are consistent with the
prediction of the Standard Model for the upper bound of \dms\ $\sim
25$~\ips\ \cite{CKM01}.\\
The work presented here gives an updated
estimate for the $\dms$ sensitivity from the ATLAS experiment using Monte Carlo events of the $\Bs$ hadronic channels $\Bs \to \Dsm\:\pi^{+}$ and $\Bs \to \Dsm\:a_{1}^{+}$ with $\Dsm \to \phi(K^{+}\,K^{-})\:\pi^{-}$ and $a_{1}^{+} \to \pi^{+}\:\pi^{-}\:\pi^{+}$.
the following sections contain a brief
discussion of the event selection, analysis cuts and the most important
kinematic distributions of the $\Bs$ candidates.

\section{\label{tools}Tools and detector layout}
A detailed description of the generation, simulation, reconstruction and analysis software tools used for this study as well as a short characterization of the properties of the ATLAS Inner Detector layout is given in \cite{BsMixRome}. The ATLAS B-physics trigger with various strategies for B-trigger selections is described in \cite{BTrig}.

\section{\label{ana}Event selection and analysis results}
\begin{figure}[h]
\includegraphics[width=65mm]{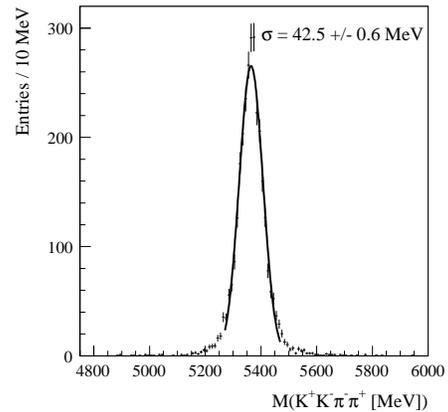}
\caption{Reconstructed $\Bs$\ invariant mass distribution
normalized to $10~\ifb$. The core of the distribution is fitted with a single Gauss function.}
\label{massbs}
\end{figure}

In the offline analysis the $\Bs$ meson is reconstructed from its decay products, applying kinematical cuts on tracks, kinematical and mass cuts on intermediate particles like $\Dsm$ and $\phi$. A vertex fit includes mass constraints and requires that the
total momentum of the $\Bs$~vertex points to the primary vertex and the
momentum of the $\Dsm$~vertex points to the $\Bs$~vertex. To improve the purity of the sample cuts on properties of the $\Bs$ candidates like proper time, impact parameter, transverse momentum and mass cuts are imposed. For the $\Bs \to \Dsm\:\pi^{+}$ channel Figure~\ref{massbs} shows the fitted invariant mass distribution $M_{KK\pi\pi}$ with a mass resolution of $\sigma_{B_s}$~=~42.5~MeV (single Gauss fit).

The proper time of the reconstructed $\Bs$~candidates is computed from the reconstructed transverse decay length $d_{xy}$, the $\Bs$~mass and the $\Bs$~transverse momentum $\pT$. Parameterized with the sum of two Gauss functions around the same mean value the widths of the two Gaussians resulting from the fit are
$\sigma_{1}$~=~(70.3$\pm$3.9)~fs for the core fraction of 54.7\% and
$\sigma_{2}$~=~(156.1$\pm$6.8)~fs for the rest of the tail part of the
distribution.

\begin{figure}[h]
\includegraphics[width=65mm]{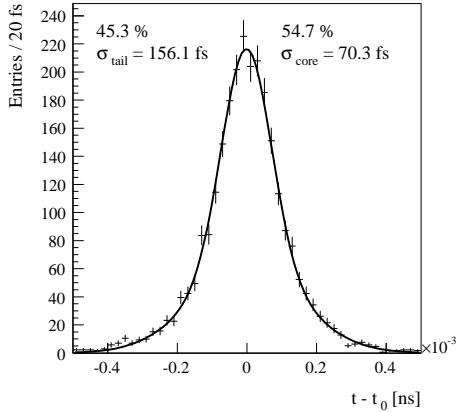}
\caption{Proper-time resolution normalized to $10~\ifb$ for $\Bs$\ from
$\Bs \to \Dsm \pip$ decays and fitted with the sum of two Gauss functions around a common mean value.}
\label{bspTime}
\end{figure}
The estimation of the maximum value of \dms\ measurable with ATLAS is
considering $\Bs$~candidates
from the $\Dsm\pip$ and $\Dsm\aonep$ channels.
The detailed analysis is done for the $\Bs \to \Dsm\:\pi^{+}$ channel, whereas numbers for the $\Bs \to \Dsm\:a_{1}^{+}$ signal and $B_{d}^{0}$ exclusive background channels are estimated extrapolations.
More details on the event selection including cut values, kinematical resolutions, expected number of events for signal and background channels can be found in \cite{BsMixRome}.

\section{Building of the likelihood function}
The probability density to observe an initial $B_{j}^{0}$ meson ($j=d,
\;s$) decaying at time $t_0$ after its creation as a $\bar{B_{j}^{0}}$
meson is given by
\begin{eqnarray}
  p_j(t_0, \; \mu_0) = \
  & \frac{\Gamma_{j}^{2} - \left( \frac{\Delta\Gamma_{j}}{2}\right)^{2}}
  {2\,\Gamma_{j}}\ \mathrm{e}^{- \Gamma_{j} t_0} \times \nonumber \\
  &
  \left ( \cosh \frac{\Delta\Gamma_{j} t_0}{2} + \mu_0 \cos\Delta m_{j} t_0
  \right)
  \label{oscPr}
\end{eqnarray}

where $\Delta\Gamma_{j} =
\Gamma_\mathrm{H}^{j}-\Gamma_\mathrm{L}^{j}$,\ $\Gamma_j =
(\Gamma_\mathrm{H}^{j} + \Gamma_\mathrm{L}^{j})/2$ and $\mu_0 = -1$. \\
For the unmixed case (an initial $B_{j}^{0}$ meson decaying as a $B_{j}^{0}$ meson at time $t_0$), the probability density is obtained by setting $\mu_0 = +1$ in Eq.~\ref{oscPr}.\\
Experimental effects like the wrong tag fraction and the resolution of the reconstructed proper time of the $\Bs$ modify the probability function. Exclusive oscillating background channels and non-oscillating combinatorial background are taken into account as fractions of the probability density called pdf$_{k}$. \\
The likelihood of the total sample is written as
\begin{equation} {\cal L}(\Delta m_s, \Delta \Gamma_s) =
  \prod_{k=1}^{N_{\mathrm{ch}}^{ }} \; \prod_{i=1}^{N_{\mathrm{ev}}^k}
  \mathrm{pdf}_k(t_i,\mu_i)
  \label{LL}
\end{equation}
The index $k=1$ denotes the $\Bs \to \Dsm \pip$ channel and $k=2$ the
$\Bs \to \Dsm \aonep$ channel,
$N_{\mathrm{ev}}^k$ is the total number of events of type $k$,
and $N_{\mathrm{ch}} = 2$. See \cite{BsMixRome} for detailed information on the building of the probability density functions. \\

\section{Creation of the \boldmath{$\Bs$}\ `data sample'}
A simplified Monte Carlo method is applied to generate a $\Bs$~sample using
the numbers of reconstructed $\Bs$ events and kinematic distributions obtained from the simulation studies in Ref. \cite{BsMixRome} as input parameters.
$\Bs$ signal events
oscillating with a given frequency \dms\
(e.g. $\dms = 100~\ips$, which is far off the expected value for \dms),
together with
$N_{\Bd}=N_{\Bd}^1 + N_{\Bd}^2$ background events oscillating with
frequency \dmd\ and $N_\mathrm{cb}=N_\mathrm{cb}^1+N_\mathrm{cb}^2$
combinatorial events (no oscillations) are generated according to
Eq.~\ref{oscPr}.

\section{Results on \boldmath{$\dms$} measurement limits}
The \dms\ measurement limits are obtained applying the amplitude fit
method~\cite{AFit} to the `data sample' generated as described in the
previous section. According to this method a new parameter, the
$\Bs$~oscillation amplitude ${\cal A}$, is introduced in the
likelihood function by replacing the term `$\mu_0 \cos\Delta m_{s}
t_0$' with `$\mu_0 {\cal A} \cos\Delta m_{s} t_0$' in the
$\Bs$~probability density function given in Eq.~\ref{oscPr}.  For each
value of \dms, the new likelihood function is minimized with respect
to ${\cal A}$, keeping all other parameters fixed, and a value ${\cal
  A} \pm \sigma_{\cal A}^{\mathrm{stat}} $ is obtained.  One expects,
within the estimated uncertainty, ${\cal A} = 1$ for \dms\ close to
its true value, and ${\cal A} = 0$ for \dms\ far from the true value.
A $5\;\sigma$ measurement limit is defined as the value of \dms\ for
which $1/\sigma_{\cal A} =5$, and a sensitivity at 95\% confidence
level as the value of \dms\ for which $1/\sigma_{\cal A} = 1.645$.
Limits are computed with the statistical uncertainty $\sigma_{\cal
  A}^{\mathrm{stat}}$. A detailed investigation on the systematic
uncertainties $\sigma_{\cal A}^{\mathrm{syst}}$, which affects the
measurement of the $\Bs$~oscillation, is presented in
\cite{BsMix}.\\
For the nominal set of parameters (as defined in the previous
sections), $\dgs = 0$ and an integrated luminosity of 30~\ifb\ the
amplitude $\pm 1 \sigma_{\cal A}^{\mathrm{stat}}$ is plotted as a
function of \dms\ in Fig.~\ref{ampl_dms}. The 95\% CL sensitivity to
measure \dms\ is found to be 30.5 $\ips$. This value is given by the
intersection of the dashed line, corresponding to $1.645\ \sigma_{\cal
  A}^{\mathrm{stat}}$ with the ${\cal A} = 1$ horizontal line.\\

\begin{figure}[h]
\includegraphics[width=60mm]{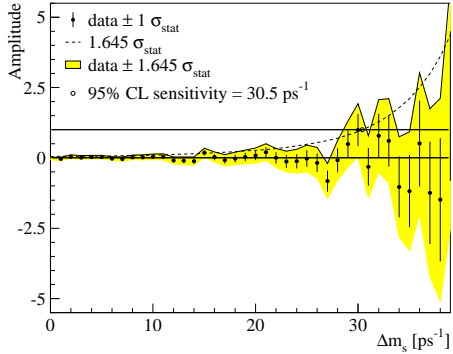}
\caption{The $\Bs$\ oscillation amplitude as a function of
$\dms$\ for an integrated luminosity of 30~\ifb\ for a specific
Monte Carlo experiment.}
\label{ampl_dms}
\end{figure}

From Fig.~\ref{sign_dms}, which shows the significance of the
measurement $S(\dms) = 1/ \sigma_{\cal A}$ as a function of \dms, the
$5\sigma$ measurement limit is found to be
22 $\ips$.
\begin{figure}[h]
\includegraphics[width=60mm]{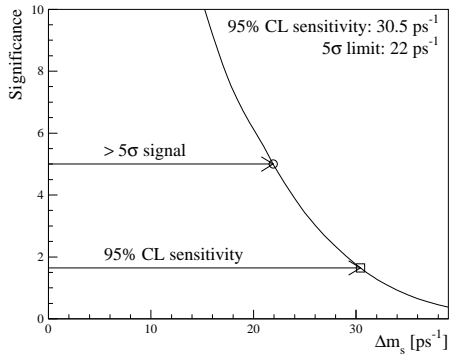}
\caption{The measurement significance as a function of
$\dms$\ for an integrated luminosity of $30~\ifb$.}
\label{sign_dms}
\end{figure}

\begin{table}[h]
  \begin{center}
      \begin{tabular}{ccc}
        \hline
        Lumi       & $5 \sigma$ limit         &   $95$\% CL sensitiv. \\
        ($\ifb$)     &   ($\ips$)                 &          ($\ips$)       \\
        \hline
        \phantom{1}5 &     13.2               &      23.8              \\
        10 &     16.5                 &      26.5                \\
        20 &     20.0                 &      29.0                \\
        30 &     21.9                 &      30.5                \\
        \hline
      \end{tabular}
    \caption{The dependence of $\dms$ measurement limits on the
      integrated luminosity.}
      \label{tab_dms_lum}
  \end{center}
\end{table}

The dependence of the \dms\ measurement limits on the integrated
luminosity is shown in Fig.~\ref{dms_lum}, with the numerical values
given in Table~\ref{tab_dms_lum}.

\begin{figure}[h]
\includegraphics[width=65mm]{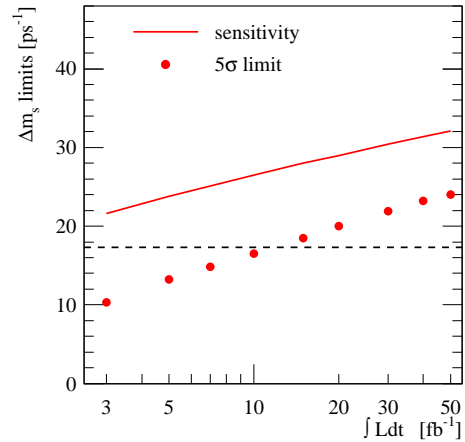}
\caption{The dependence of $\dms$\ measurement limits on
the integrated luminosity. The dotted horizontal line gives the
CDF measurement \cite{CDF}.}
\label{dms_lum}
\end{figure}

The dependence of the \dms\ measurement limits on \dgsg\ is determined
for an integrated luminosity of 30~\ifb, other parameters having their
nominal value. The \dgsg\ is used as a fixed parameter in the
amplitude fit method. As shown in Fig.~\ref{dms_dg} no sizeable effect
is seen up to a $\dgsg$ of 50\%.\\

\begin{figure}[h]
\includegraphics[width=65mm]{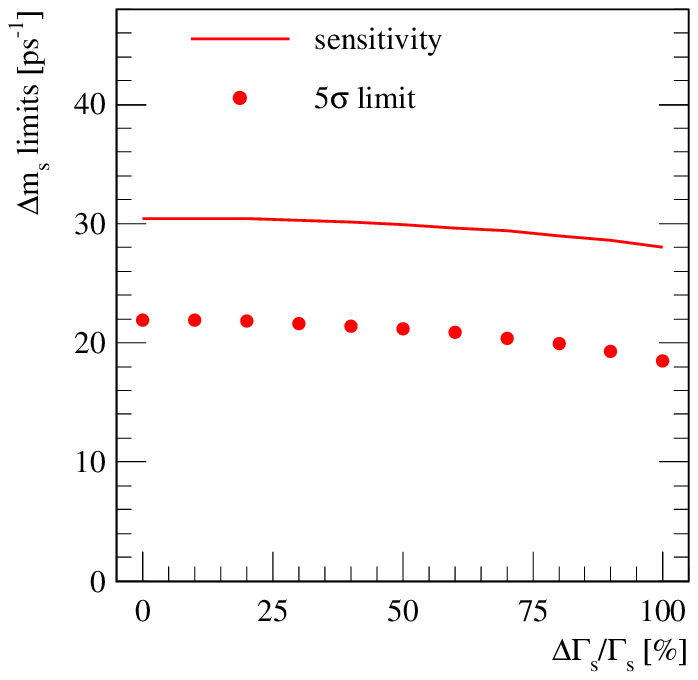}
\caption{The dependence of \dms\ measurement limits on
\dgsg.}
\label{dms_dg}
\end{figure}

\newpage
\section{\label{conclusion}Conclusions}
In this summary the performance of the $\Bs \to
\Dsm(\phi\,\pi^{-})\:\pi^{+}$ channel and extrapolated numbers for $\Bs \to \Dsm\:a_{1}^{+}$ channel are used to calculate the 95\% CL
exclusion and $5\;\sigma$ measurement limits of the \Bs\ oscillation
frequency as a function of the integrated luminosity collected with
the ATLAS detector.  The limits are updated for the detector geometry
of ``initial layout'' using full Rome statistics, but only statistical
errors are taken into account.  With an integrated luminosity from
10 to 20~\ifb\ a $5\;\sigma$ measurement for a range of $16.5 < \dms\ <
20~\ips$ is possible, covering
the recent results from the Tevatron collider. \\
The values obtained in this note for the measurement limits should be
re-evaluated, taking into account changes in the detector geometry,
especially ``complete detector layout'', and the evolving simulation
and reconstruction software.  $\Dsm\aonep$ and exclusive \Bd\
background channels will be analyzed independently and investigations
looking at the performance of other interesting $\Bs$-$\barBs$ mixing
channels, which might be included in the analysis, will be carried
out.

\bigskip 
\begin{acknowledgments}

Work supported by Bundesministerium fuer Bildung, Wissenschaft und Kultur, Austria.
\end{acknowledgments}

\bigskip 

\end{document}
%